\documentclass[fleqn,usenatbib]{mnras}
\usepackage{hyperref}
\usepackage{newtxtext,newtxmath}

\usepackage[T1]{fontenc}
\usepackage{xcolor}

\DeclareRobustCommand{\VAN}[3]{#2}
\let\VANthebibliography\thebibliography
\def\thebibliography{\DeclareRobustCommand{\VAN}[3]{##3}\VANthebibliography}

\newcommand{\codefont}[1]{{\texttt{#1}}}
\usepackage{graphicx}	
\usepackage{amsmath}	
\usepackage{float}
\usepackage{accents}
\usepackage{cancel}

\title[$H_0$ inference from simulated SNeIa]{Field-level inference of $H_0$ from simulated type Ia supernovae in a local Universe analogue}

\author[E. Tsaprazi \& A. F. Heavens]{
Eleni Tsaprazi$^{1}$,\thanks{e.tsaprazi@imperial.ac.uk}
Alan F. Heavens$^{1}$\\
$^{1}$Imperial Centre for Inference and Cosmology (ICIC) \& Astrophysics group, Department of Physics, Imperial College, Blackett Laboratory,\\ Prince Consort Road, London SW7 2AZ, UK\\
}

\date{Accepted XXX. Received YYY; in original form ZZZ}

\pubyear{2025}

\begin{document}
\label{firstpage}
\pagerange{\pageref{firstpage}--\pageref{lastpage}}
\maketitle

\begin{abstract}
Two particular challenges face type Ia supernovae (SNeIa) as probes of the expansion rate of the Universe. One is that they may not be fair tracers of the matter velocity field, and the second is that their peculiar velocities distort the Hubble expansion. Although the latter has been estimated at $\lesssim1.5\%$ for $z>0.023$, this is based either on constrained linear or unconstrained (random) non-linear velocity simulations. In this paper, we address both challenges by incorporating a physical model for the locations of supernovae, and develop a Bayesian Hierarchical Model that accounts for non-linear peculiar velocities in our local Universe, inferred from a Bayesian analysis of the 2M++ spectroscopic galaxy catalogue. With simulated data, the model recovers the ground truth value of the Hubble constant $H_0$ in the presence of peculiar velocities including their correlated uncertainties arising from the Bayesian inference, opening up the potential of including lower redshift SNeIa to measure $H_0$. Ignoring peculiar velocities, the inferred $H_0$ increases minimally by $\sim 0.4 \pm 0.5$ km s$^{-1}$ Mpc$^{-1}$  in the range $0.023<z<0.046$.
We conclude it is unlikely that the $H_0$ tension originates in unaccounted-for non-linear velocity dynamics. 
\end{abstract}




\section{Introduction}\label{sec:intro}
The emergence of data-constrained cosmological simulations equipped with galaxy formation models enables the study of supernova (SN) physics in realistic reconstructions of the local large-scale structure for the first time \citep[e.g.][]{2011MNRAS.417.1434F,2014MNRAS.441.2593N,2021MNRAS.504.2998S}. Unlike random N-body simulations, constrained ones enable the reconstruction of the large-scale structure from initial conditions that reproduce the observed clustering of real galaxies in the Universe. Here, we simulate SNeIa from individual galaxy properties in such an analogue of the local Universe. We demonstrate a Bayesian hierarchical framework that enables the study of the impact of non-linear peculiar velocities on $H_0$ and the inference of $H_0$ from SNeIa at $z<0.023$, which are typically discarded. The latter is enabled by the fact that realistic peculiar velocity reconstructions in the local Universe are available.

We build upon the \codefont{SIBELIUS-DARK} \citep{SIBELIUS,SIBELIUS-DARK} cosmological simulation, whose initial conditions were constrained with 2M++ galaxies \citep{2011MNRAS.416.2840L} through the \codefont{BORG} algorithm \citep{2013MNRAS.432..894J,2016MNRAS.455.3169L,2019A&A...625A..64J}, out to a distance of $200$ Mpc ($z=0.046$). Whilst \codefont{SIBELIUS-DARK} is a dark-matter only cosmological simulation, it is equipped with \codefont{GALFORM} \citep{GALFORM}, a semi-analytic galaxy formation and evolution model that emulates baryonic physics. The \codefont{SIBELIUS-DARK} simulation provides a catalogue of simulated galaxies, each with complete star-formation histories, which we use to derive SN rates.

SNe are clustered in the cosmic large-scale structure \citep{2008ApJ...682L..25C,2008MNRAS.383.1121M,2022MNRAS.510..366T} as they typically reside in star-forming galaxies. The latter are less clustered than typical galaxies, therefore the clustering pattern of SNe may differ from such galaxies \citep{2021MNRAS.500.1071M}. At the same time, more complex intergalactic physics affects the emergence of SNe. Earlier studies assumed the distribution of SNe on the sky to be uniform \citep{2002A&A...392..757G,2019JCAP...10..005F}, whereas more recent ones simulate type Ia supernovae in individual haloes \citep{2023A&A...674A.197C}, or from galaxy properties \citep{2021MNRAS.505.2819V,2023MNRAS.520.2887L}. 
The catalogues presented here contain SNe simulated from individual galaxy properties and distributed in a realistic large-scale structure.  

We first employ the mock SN catalogues we generate to study the effect of peculiar velocities induced by our local large-scale structure on the derived Hubble constant, $H_0$, inferred from the measured redshifts and distance moduli of simulated SNeIa. The argument that local inhomogeneities affect the measured expansion dates back to \citet{1998AJ....116.1009R}. Ever since, a large number of studies have investigated the effect providing evidence in favour or against a special configuration of the cosmic large-scale structure that could result in the emergence of a discrepancy \citep{R16, R22} between measurements of $H_0$ in the local and distant Universe \citep[e.g.][]{2014MNRAS.438.1805W, 2015MNRAS.450..317C,odderskov,2019ApJ...875..145K,2020A&A...633A..19B,2020PhRvD.102b3520K,2021MNRAS.500.3728S,2022ApJ...938..112P,2022JCAP...07..003C,2022CQGra..39r4001C,2022PASA...39...46C,2023PDU....4201348P,2023PhRvD.108l3533M,2024JCAP...01..071G,carreres2024ztf,2025arXiv250115704H}, even in the absence of a local void \citep{2019A&A...625A..64J}. 

Our study adds to the above investigations by a) considering our specific large-scale structure instead of random N-body simulations which provides us with knowledge of peculiar velocities, b) using non-linear peculiar velocities at the locations of simulated SNe along with their non-linear correlations through a constrained gravity model, c) exploiting correlations with the full large-scale structure, instead of individual structures that gravitationally influence the local Universe dynamics, d) looking at the locations of SNeIa generated according to galaxy evolution. Further, since \codefont{SIBELIUS-DARK} emulates an analogue of our Local Group, our results naturally account for effects of our specific environment on the observer and therefore, potential deviations from the Copernican Principle \citep{2023A&A...671A..68C}. The demonstration presented in this study paves the way for a future analysis of observed SNIa data at $z<0.023$.

This paper is structured as follows: In Section \ref{method}, we describe the SNIa rate model and the Bayesian Hierarchical Model for the inference of $H_0$. In Section \ref{results}, we present our findings on the effect of peculiar velocities on $H_0$ measurements and further validate the model on simulated SNIa data with assumed observational uncertainties at $z<0.023$. Finally, in Section \ref{conclusions} we summarise our conclusions and provide an outlook on upcoming developments.

\begin{figure}
\centering
\includegraphics[width=0.45\textwidth]{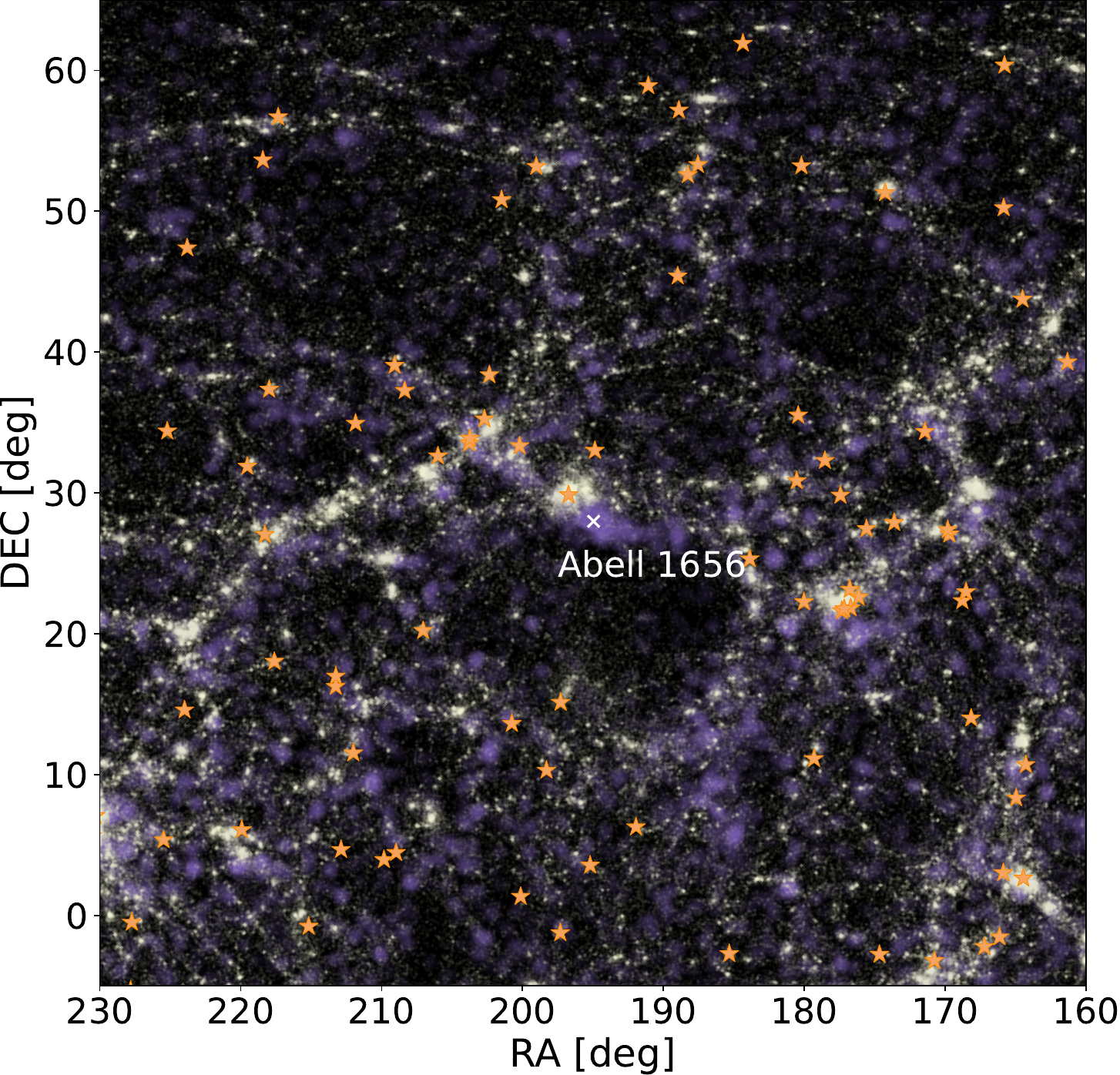}
\caption{Angular coordinates of the 2M++ galaxies (purple), simulated \codefont{SIBELIUS} galaxies (yellow) and SNeIa (orange). The Coma cluster (Abell 1656) is indicated by a white cross. The \codefont{SIBELIUS-DARK} galaxies follow the distribution of observed galaxies in the real Universe on large scales.}
\label{fig_00}
\end{figure}

\section{Method}\label{method}

Our analysis is based on the \codefont{SIBELIUS-DARK} catalogue of simulated galaxies \citep{SIBELIUS-DARK}. The catalogue was generated within the dark matter density and velocity field of the \codefont{SIBELIUS} dark-matter only simulation \citep{SIBELIUS} with the application of the \codefont{GALFORM} galaxy formation and evolution model \citep{GALFORM}. The \codefont{SIBELIUS-DARK} catalogue contains $4,527,471$ hosts with stellar masses $M>10^7 M_\odot$ at $z=0$. This mass range includes all SNIa host galaxies \citep[][Figure 1]{2023MNRAS.526.5292T}. The \codefont{GALFORM} star formation histories are informed about galaxy mergers, cooling, feedback, photoionisation, as well as the evolution of cosmic metallicity and mass-metallicity correlations. The gravity model through which the constrained initial conditions of the \codefont{SIBELIUS} simulation were inferred, further gives us access to non-linear peculiar velocities at the field level (3D reconstruction). Since \codefont{SIBELIUS} \citep{SIBELIUS} is a simulation constrained with the initial conditions inferred from the 2M++ catalogue \citep{2019A&A...625A..64J}, our analysis carries over all the conclusions reported therein. Specifically, the 2M++ large-scale structure reconstruction we use shows no evidence of a large-scale underdensity which could account for the Hubble tension \citep{2019A&A...625A..64J}. Therefore our conclusions below are not driven by a local void which distorts the observed cosmic expansion. Below, we describe our modelling of SN occurrence in each simulated galaxy.

\subsection{SNIa rate modelling}

SNIa rates, $R_\mathrm{IA}$, are typically modeled as \citep[e.g.][]{2006ApJ...648..868S,Graur,DES}
\begin{equation}
    R_\mathrm{IA} = \int_{t_0}^{t_\mathrm{f}}\Psi(t_0-\tau)\Phi(\tau)d\tau,
    \label{eq:rates}
\end{equation}
where $\tau$ is the time following a star-formation burst, $t_0$ corresponds to the lookback time to a galaxy's redshift, $t_\mathrm{f}$ indicates the lookback time to the epoch at which the first star formation burst occurred, $\Psi$ is the star formation history and $\Phi$ the delay time distribution. The latter indicates the time between the formation of a stellar population and the explosion of the first SNeIa \citep{DTD}. It can be modelled as a power-law \citep{DES}
\begin{equation}
    \Phi(\tau) = 
    \begin{cases} 0 & \text{if } \tau < t_\mathrm{p} \\
    A\left(\frac{\tau}{\mathrm{Gyr}}\right)^\beta                                  & \text{if } \tau \geq t_\mathrm{p},     %
    \end{cases}
\end{equation}
where $t_\mathrm{p}$ is the time required for a population of white dwarves to form after a star formation burst. We take $t_\mathrm{f}$ to be the lookback time to the redshift at which $\Psi$ becomes non-zero for the first time in each galaxy. $\Psi$ is provided by \codefont{SIBELIUS-DARK} for every galaxy in the simulation volume out to $z=20$. We adopt $A = 2.11 \times 10^{-13}$ SNe $M_\odot^{-1}$ yr$^{-1}$, $\beta = -1.13$, $t_\mathrm{p} = 40$ Myr \citep{DES}. We rescale all rates by $\sim3$ to match the mean global volumetric rate reported by \citet{2020ApJ...904...35P}: (2.35 $\pm$ 0.24) $\times 10^4$ SNeIa yr$^{-1}$Gpc$^{-3}$. The need for rescaling likely arises due to a combination of factors. \codefont{GALFORM} was calibrated using datasets other than SNIa ones, therefore a deviation from observed rates is in principle expected. Further, a mild underestimation of mass was reported in \codefont{SIBELIUS-DARK} in the range $\sim5\times10^9-10^{11}M_\odot$, where most SNIa hosts reside, which explains the need to apply a rescaling factor $>1$ to match the observed SNIa production efficiency.

The underestimation of the global volumetric rate affects almost the entire population systematically, so the relative object-by-object weights we recover, which are of interest in our analysis, are credible. Such rescalings have been performed in hydrodynamical simulations when comparing the derived SNIa rates to observations \citep[Figure 3,][]{EAGLE}. Our analysis is by construction insensitive to global rate rescalings, since it is only affected by the SNIa host velocity and distance distribution. As we will demonstrate in Section \ref{results}, it is also on average insensitive to the per-host rate modelling given the local Universe realisation. It is the subject of future work to test various SNIa rate and galaxy formation models against observations, since it has been shown that the assumptions on the star-formation history may significantly affect astrophysical predictions \citep{2022MNRAS.514.1315B}.

Finally, we model the occurrence of SNe in a galaxy in a given period of time $\Delta t$, as a Poisson process \citep{DES}
\begin{equation}
    \mathcal{P}(N_\mathrm{s}|R_\mathrm{IA}) = \frac{\lambda_\mathrm{SN}^{N_\mathrm{s}} e^{-\lambda_\mathrm{SN}}}{N_\mathrm{s}!},
    \label{eq:Poisson}
\end{equation}
where $N_\mathrm{s}$ is the number of SN explosions per galaxy, and $\lambda_\mathrm{SN} = R_\mathrm{IA} \Delta t$ the intensity of the Poisson process. In order to obtain the SN mock catalogues, we sample from the above distribution. 

\subsection{Hubble constant inference}

We will use realisations drawn from Equation \ref{eq:Poisson} to validate a framework for the incorporation of peculiar velocities in inference of $H_0$ and quantify the effect of our local velocity environment on $H_0$. The latter impact is given by the fractional Hubble uncertainty
\begin{equation}
    \frac{\Delta H}{H_0} = \frac{H_0-H_{0, \mathrm{fid}}}{H_{0, \mathrm{fid}}}
    \label{eq:var1}
\end{equation}
where $H_{0,\mathrm{fid}}$ represents a fiducial value and $H_0$ the Hubble constant inferred if peculiar velocities are properly accounted for. The fractional uncertainty can also be written as the intercept of the Hubble diagram, $\alpha_\mathrm{B}$ \citep{R22,2023ApJ...954L..31S,carreres2024ztf}

\begin{equation}
    \log{H_{0,\mathrm{fid}}} = \frac{M_\mathrm{B}+5\alpha_\mathrm{B, fid}+25}{5},
    \label{eq:var2}
\end{equation}
where $M_\mathrm{B}$ is the absolute magnitude of SNeIa in the B-band. We write Equation \ref{eq:var2} once for the fiducial and once for the inferred Hubble constant and subtract the former from the latter, assuming that $M_\mathrm{B}$ is fixed. Recasting the resulting terms into the fractional Hubble variation with the help of Equation \ref{eq:var1}, we arrive at a proxy for the fractional variation in the Hubble constant
\begin{equation}
    \alpha_\mathrm{B} =  \mu-\mu_\mathrm{fid} = \log{\left(\frac{\Delta H}{H_0}+1\right)},
    \label{eq:transform}
\end{equation}
which we will use with $\Delta H/H_0$ interchangeably according to the above equation, since it represents the difference of the inferred from the fiducial Hubble constant. While the simulation extends in $0<z<0.046$, we will truncate it to $z>0.023$ in which the magnitude-redshift relation is computed for the purposes of investigating the impact of peculiar velocities on $H_0$ \citep{2021MNRAS.505.3866E}. We will also consider SNeIa at $z<0.023$ to demonstrate that our framework allows us to account for peculiar velocities for SNeIa which are typically discarded from $H_0$ analyses. This is the only redshift cut applied for part of our analysis, for comparison with typical cuts made in the literature to avoid excessive peculiar velocity corrections. The \codefont{SIBELIUS-DARK} catalogue is otherwise redshift complete. Note that the Bayesian hierarchical model that we present would allow lower redshift SNeIa to be used in an analysis of $H_0$.

\subsection{Non-linear velocity covariance matrix estimation}

Here, we use our knowledge of the non-linear peculiar velocity field from the 2M++ Bayesian reconstruction \citep{2019A&A...625A..64J} to reduce the impact of peculiar velocities. The Bayesian reconstruction provided a posterior of the initial conditions for structure formation which were evolved to present-day density and velocity fields that reproduce the observed clustering of the 2M++ spectroscopic galaxy catalogue. The velocity fields were estimated using a Cloud-in-Cell (CiC) kernel \citep[e.g.][]{hockney2021computer,2021A&A...646A..65M}. The minimum scale at which we trust the physical modelling of peculiar velocities in the 2M++ reconstruction is 2.65 Mpc/h. For a real data application, the velocity dispersion on smaller scales must be modelled externally. 

\citet{carreres2024ztf} argued that linear peculiar velocity correlations can affect the derived $\alpha_\mathrm{B}$. We take into account both the non-linear velocity correlations and the varying quality of the 2M++ reconstruction as a function of location in the reconstruction box by accounting for the full non-linear peculiar velocity covariance matrix. We consider $N_\mathrm{sim} \sim 240$ non-linear velocity fields of the 2M++ reconstruction at the locations of simulated SNeIa for each realisation of supernova positions \citep{2019A&A...625A..64J}. This provides us with $N_\mathrm{sim} \times N_\mathrm{SN}$ velocity values from which we build an $N_\mathrm{SN} \times N_\mathrm{SN}$ covariance matrix per SNIa dataset, which we denote by $C_\mathrm{2M++}$ in what follows. The covariance matrix is then estimated as
\begin{equation}
C_\mathrm{2M++} = \frac{1}{N_\mathrm{sim} - 1} \sum_{i=1}^{N_\mathrm{sim}} \left( v^{i} - \bar{v} \right) \left( v^{i} - \bar{v} \right)^\top,
\end{equation}
where $v^i$ is the velocity field at the locations of SNe in realisation $i$ and $\bar{v}$ is the mean velocity field across all $N_\mathrm{sim}$ realisations at the SN locations. The number of velocity fields, $N_\mathrm{sim}$, is determined by the data availability and is statistically adequate for the purposes of our analysis, because within that number, all uncertainties in the reconstruction have been captured by the velocity covariance matrices. Our covariance matrices are robust to this number. The velocity fields are created by forward-modelling from initial conditions that are sampled from the \codefont{BORG} posterior. The variability in the samples arises from various sources, such as Poisson shot noise of galaxies and prior ranges for parameters. At the same time, each velocity field, having been constructed with a gravity model, accounts for physical correlations between any points. We choose to simulate $N_\mathrm{SN}$ SNe, such that $N_\mathrm{SN} < N_\mathrm{sim}-2$, otherwise the estimated covariance matrix is not invertible. We generate multiple SNIa Poisson samples which satisfy the above condition. We keep only those that contain at most 1 SN per voxel, such that information is not repeated in the covariance matrix. These two conditions guarantee that we end up with invertible covariance matrices for each SN realisation. In our setting, these conditions are satisfied for $N_\mathrm{SN} \sim 200$ SNe (within Poisson noise) per dataset.

The impact of peculiar velocities on $H_0$ in observations is typically inferred from distance modulus $\hat{\mu}$ and redshift estimates $\hat{z}$, which are known with some noise $\sigma_\mathrm{\mu}$ and $\sigma_\mathrm{z}$, respectively. The error on distance moduli can be significant and therefore cannot simply be added in quadrature to redshift uncertainties as is typically done within the errors assumed here. More importantly, at $z<0.023$ where the contribution of peculiar velocities to the observed redshift is significant, the dependence of distance moduli on the peculiar velocities must be self-consistently modelled along each line of sight. In this section, we advance the formalism in \citet{2017JCAP...10..036R} to infer the intercept of the Hubble diagram from redshift and distance moduli data, by considering their uncertainties along both their corresponding axes, instead of adding them in quadrature. This requires a Bayesian Hierarchical Model that accounts for the known non-linear 2M++ peculiar velocities in $H_0$ inference.

\begin{figure}
\centering
\includegraphics[width=0.45\textwidth]{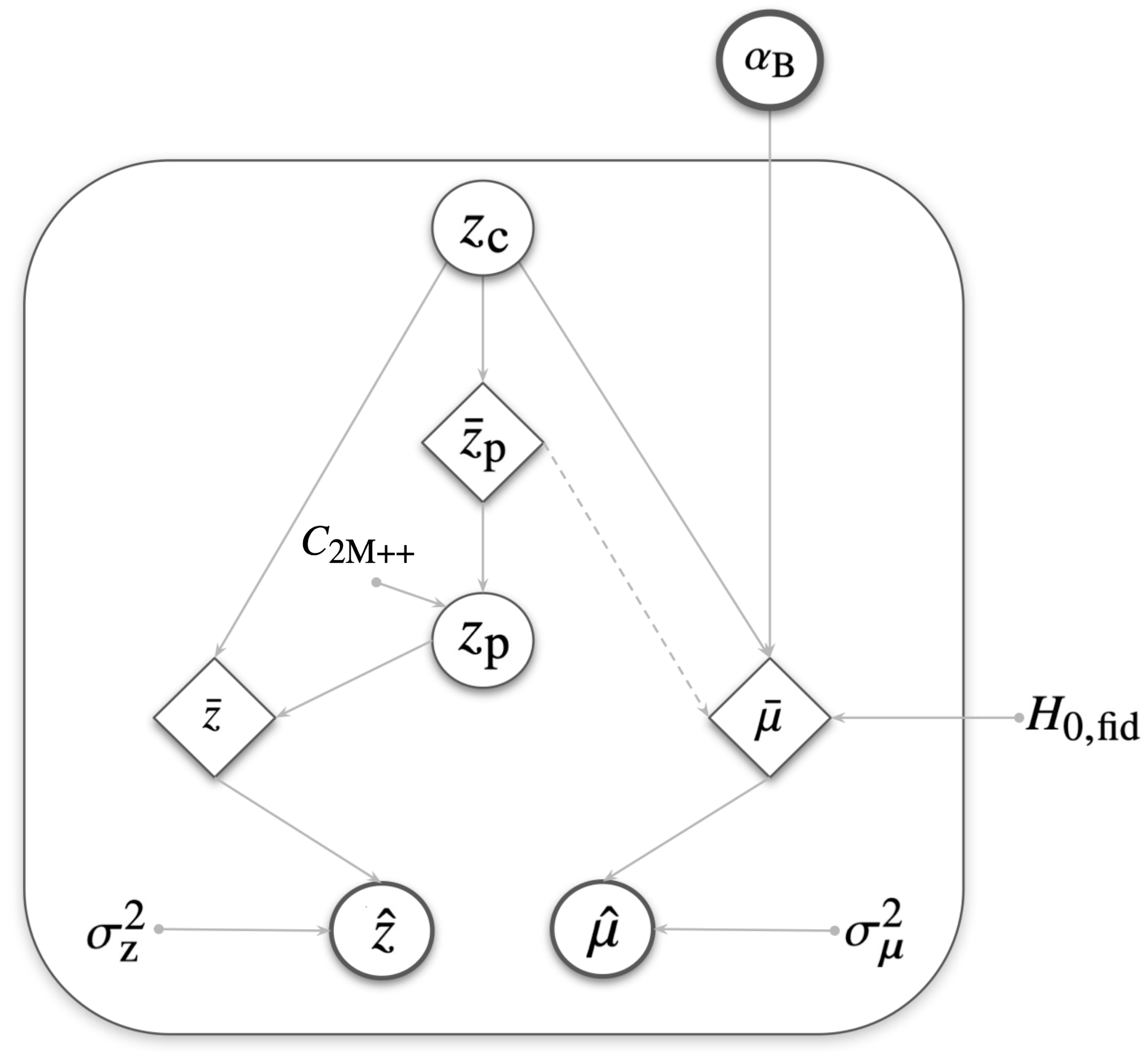}
\caption{Flowchart of the the Bayesian hierarchical model in Equation \ref{eq:POSTERIOR}. The rhombi represent the means of the multivariate likelihoods. The thin circles represent the stochastic variables. The bold circle represents the population parameter $\alpha_\mathrm{B}$. At any given cosmological redshift, we use the 2M++ non-linear velocity reconstruction to incorporate the effect of redshift-space distortions and the stochasticity introduced by the non-linear covariance matrix and assumed observational uncertainties. The dashed line indicates a necessary approximation for the efficient computation of the posterior.}
\label{fig_2}
\end{figure}

\subsection{Hubble constant inference}\label{sec:H0inf}

We will consider a set of observed redshifts, $\hat{z}$, observed distance moduli, $\hat{\mu}$, and the three-dimensional reconstruction of velocities from which we obtain samples of peculiar redshifts at any cosmological redshift, $z_\mathrm{c}$. These data are given in the presence of uncertainties which render the exact true value of peculiar redshifts, ${z}_\mathrm{p}$, and cosmological redshifts unknown. We characterise the posterior peculiar redshifts with means $\bar{z}_\mathrm{p}$ and covariance $C_{2M++}$ from the 2M++ reconstruction. As all quantities except for $\alpha_\mathrm{B}$ are vectors, we will omit writing them in bold. Marginalising over $z_\mathrm{p}$ and $z_\mathrm{c}$, the posterior may be written as (note that this is a multidimensional integral over all the supernova redshifts)
\begin{eqnarray}
p(\alpha_\mathrm{B}|\hat{z}, \hat{\mu}, \bar{z}_\mathrm{p}) &\propto & p(\hat{z}, \hat{\mu}|\alpha_\mathrm{B}, \bar{z}_\mathrm{p})p(\alpha_\mathrm{B}|\cancel{\bar{z}_\mathrm{p}})\\
&\propto& \pi(\alpha_\mathrm{B})\int p(\hat{z},\hat{\mu},z_\mathrm{p},z_\mathrm{c}|\alpha_\mathrm{B},\bar{z}_\mathrm{p})dz_\mathrm{p}dz_\mathrm{c}\\
&\propto& \pi(\alpha_\mathrm{B})\int p(\hat{\mu}|\cancel{\hat{z}},z_\mathrm{p},z_\mathrm{c}, \alpha_\mathrm{B},\bar{z}_\mathrm{p})\nonumber\\
&\times&p(\hat{z},z_\mathrm{p},z_\mathrm{c}|\alpha_\mathrm{B},\bar{z}_\mathrm{p})dz_\mathrm{p}dz_\mathrm{c},
\label{eq:initial_posterior}
\end{eqnarray}
where $ \pi(\alpha_\mathrm{B})$ is the prior on $\alpha_\mathrm{B}$ (which we will take to be uniform), and we have assumed that the estimated redshifts and distance moduli conditioned on the true values are independent. We have removed variables where there is no dependence with $\cancel{\phantom.}$. The distance modulus is given by $\bar{\mu}=\mu_\mathrm{c}+10\log(1+z_\mathrm{p})-5\alpha_\mathrm{B}$, $\mu_\mathrm{c}$ being the fiducial distance modulus at any given cosmological redshift $z_\mathrm{c}$. The Doppler boosting term is a small correction, and its dependence on the true $z_\mathrm{p}$ complicates the integral enormously and prevents a necessary simplification, so we approximate this dependence by replacing $z_\mathrm{p}$ by $\bar{z}_\mathrm{p}$, i.e. we take
\begin{equation}
\bar{\mu}\simeq\mu_\mathrm{c}(H_{0,\mathrm{fid}}, z_\mathrm{c})+10\log[1+\bar{z}_\mathrm{p}(z_\mathrm{c})]-5\alpha_\mathrm{B}.
\end{equation}
This assumption has very little impact on our analysis, as our final posterior on $H_0$ is insensitive to it. We simplify the posterior in Equation \ref{eq:initial_posterior}
\begin{eqnarray}
p(\alpha_\mathrm{B}|\hat{z}, \hat{\mu}, \bar{z}_\mathrm{p}) &\propto &\pi(\alpha_\mathrm{B}) \int dz_\mathrm{p}dz_\mathrm{c}\, p(\hat{\mu}|z_\mathrm{c},\bar{z}_\mathrm{p}, \alpha_\mathrm{B})\nonumber\\
&\times&p(\hat{z}|z_\mathrm{p},z_\mathrm{c},\cancel{\alpha_\mathrm{B}},\cancel{\bar{z}_\mathrm{p}})p(z_\mathrm{p},z_\mathrm{c}|\alpha_\mathrm{B},\bar{z}_\mathrm{p}).
\end{eqnarray}
We also neglect a small correction due to selection on the estimated redshifts $\hat z$. Retaining only the relevant dependencies, the integral simplifies to
\begin{eqnarray}
p(\alpha_\mathrm{B}|\hat{z}, \hat{\mu}, \bar{z}_\mathrm{p}) &\propto &\pi(\alpha_\mathrm{B}) \int  dz_\mathrm{p}dz_\mathrm{c}\, p(\hat{\mu}|z_\mathrm{c},\bar{z}_\mathrm{p},\alpha_\mathrm{B})\nonumber\\
&\times&p(\hat{z}|z_\mathrm{p},z_\mathrm{c})p(z_\mathrm{p}|\bar{z}_\mathrm{p})p(z_\mathrm{c}).
\end{eqnarray}
Reordering the integrals, we arrive at
\begin{eqnarray}
p(\alpha_\mathrm{B}|\hat{z}, \hat{\mu}, \bar{z}_\mathrm{p}) &\propto &\pi(\alpha_\mathrm{B}) \int dz_\mathrm{c}\, p(\hat{\mu}|z_\mathrm{c},\bar{z}_\mathrm{p},\alpha_\mathrm{B})p(z_\mathrm{c}) \nonumber \\
&\times&\left[\int dz_\mathrm{p}\,p(\hat{z}|z_\mathrm{p},z_\mathrm{c})p(z_\mathrm{p}|\bar{z}_\mathrm{p})\right].
\label{eq:reordering}
\end{eqnarray}

In what follows, we will formulate the integral over $z_\mathrm{p}$ analytically to ensure tractability in high-dimensional spaces. Following the derivation in \citet{2016MNRAS.456L.132S}, $p(z_\mathrm{p}|\bar{z}_\mathrm{p})$ the likelihood marginalised over the uncertain true covariance matrix is a multivariate Student-t likelihood (see also \citet{2022MNRAS.510.3207P}). However, this does not allow analytic marginalisation over $z_\mathrm{p}$, so we instead use the Hartlap-corrected Gaussian approximation with covariance $C$, which provides a debiased estimate of a covariance matrix that has been estimated with uncertainty, as the one in 2M++. The new covariance reads \citep{2007A&A...464..399H}, $
    C = C_\mathrm{2M++}{N_\mathrm{sim}}/({N_\mathrm{sim}-N_\mathrm{SN}-1})$.
The mean is $\bar{z}_\mathrm{p}$. For the redshift likelihood $p(\hat{z}|z_\mathrm{p},z_\mathrm{c})$, we will assume independent Gaussian errors with covariance $\Sigma=$ diag($\sigma_\mathrm{z}^2$), and mean $\bar{z}=(1+z_\mathrm{c})(1+z_\mathrm{p})-1$ \citep{2011ApJ...741...67D}. For shortness of notation, we define
\begin{equation}
I = \int dz_\mathrm{p}\, p(\hat{z}|z_\mathrm{p},z_\mathrm{c})p(z_\mathrm{p}|\bar{z}_\mathrm{p}),
\end{equation}
which becomes
\begin{eqnarray}
I &=& \int dz_\mathrm{p}\, \frac{1}{\sqrt{|2\pi C|}}\frac{1}{\sqrt{|2\pi \Sigma|}} \exp\left(-\frac{1}{2}q^\mathrm{T} \Sigma^{-1}q\right)\nonumber \\
&\times&\exp\left[-\frac{1}{2}(z_\mathrm{p}-\bar{z}_\mathrm{p})^\mathrm{T}C^{-1}(z_\mathrm{p}-\bar{z}_\mathrm{p})\right],
\end{eqnarray}
where
\begin{equation}
q = \hat{z}-[(1+z_\mathrm{c})(1+z_\mathrm{p}(z_\mathrm{c}))-1].
\label{eq:q}
\end{equation}
By rearranging the terms in the first term in the integrand above, we arrive at
\vspace{-1mm}
\begin{eqnarray}
I &=& \int dz_\mathrm{p}\,\frac{1}{\sqrt{|2\pi C|}}\frac{1}{\sqrt{|2\pi \Sigma|}} \nonumber \\ [-2pt]
&\times&\exp\left[-\frac{1}{2}(z_\mathrm{p}-\tilde{z})^\mathrm{T}\tilde{\Sigma}^{-1}(z_\mathrm{p}-\tilde{z})\right]\nonumber\\[-2pt]
&\times&\exp\left[-\frac{1}{2}(z_\mathrm{p}-\bar{z}_\mathrm{p})^\mathrm{T}C^{-1}(z_\mathrm{p}-\bar{z}_\mathrm{p})\right], 
\end{eqnarray}
where
\vspace{-2mm}
\begin{eqnarray}
\tilde{z} &=& (z_\mathrm{c}-\hat{z}) \times {\rm diag}({z_\mathrm{c}+1})^{-1},\\
\tilde{\Sigma} &=& \Sigma\times {\rm diag}(1+z_\mathrm{c})^{-2}.
\end{eqnarray}
Substituting $u=z_\mathrm{p}-\bar{z}_\mathrm{p}$ and $v=\tilde{z}-\bar{z}_\mathrm{p}$ above and evaluating the n-dimensional Gaussian integral gives
\begin{eqnarray}
I &=& \int du \,\frac{\sqrt{|2\pi(C^{-1}+\tilde{\Sigma}^{-1})^{-1}|}}{\sqrt{|2\pi C|}\sqrt{|2\pi \Sigma|}}\nonumber \\
&\times&\exp\left\{-\frac{1}{2}v^\mathrm{T}\left[\tilde{\Sigma}^{-1}-\tilde{\Sigma}^{-1}(C^{-1}+\tilde{\Sigma}^{-1})\tilde{\Sigma}^{-1}\right]v\right\}.
\end{eqnarray}
Using the \citet{MR38136} formula and the matrix determinant theorem we find
\begin{equation}
I = \frac{1}{\sqrt{|2\pi(C+\tilde{\Sigma})|}} \exp\left[-\frac{1}{2}v^\mathrm{T}\left(\tilde{\Sigma}+C\right)^{-1}v\right].
\end{equation}

Then, our final posterior reads
\begin{eqnarray}
p(\alpha_\mathrm{B}|\hat{z}, \hat{\mu}, \bar{z}_\mathrm{p}) &\propto &\pi(\alpha_\mathrm{B}) \int dz_\mathrm{c}\, z_\mathrm{c}^2 \nonumber \\
&\times&\exp\left[-\frac{1}{2}(\hat{\mu}-\bar{\mu})^\mathrm{T}\Sigma_\mathrm{\mu}^{-1}(\hat{\mu}-\bar{\mu})\right] \nonumber \\
&\times&\exp\left[-\frac{1}{2} \bar q^\mathrm{T}\left(\Sigma+\tilde{C}\right)^{-1}\bar q\right],
\label{eq:POSTERIOR}
\end{eqnarray}
where $\bar q = \hat{z}-[(1+z_\mathrm{c})[1+\bar{z}_\mathrm{p}(z_\mathrm{c})]-1]$, $\tilde{C} = C\times {\rm diag}(1+z_\mathrm{c})^2$, $\Sigma_\mathrm{\mu}=$ diag($\sigma_\mathrm{\mu}^2$),
and we have assumed a $z_\mathrm{c}^2$ prior on cosmological redshifts, motivated by the volume effect in a low-redshift survey. We find that deviations from $z_\mathrm{c}^2$ are minimal and have insignificant impact on our conclusions in our redshift range. This is an integral of the same dimensionality as the $\hat{\mu}$ vector. The flowchart of the method is shown in Figure \ref{fig_2}. 

In the above formulation, we have further assumed that the change in $H_0$ leaves the peculiar velocity reconstruction invariant. It is unlikely that changes in the reconstructed velocity field with $H_0$ affect our inference. In the regions of highest sensitivity of the density field to $H_0$ \citep[Figure 2, ][]{2022A&A...657L..17K}, changes of $5 $ km s$^{-1}$ Mpc$^{-1}$, would result in a change in overdensity of no more than $\Delta \delta \approx 0.25$ at the resolution of the 2M++ inference, in filaments where the density is typically $\delta \sim 6$ at this resolution \citep[Figure 9,][]{2019A&A...625A..64J}, so the effect is likely to be small. Moreover the peculiar velocity field has a longer correlation length than the dark matter density field, and outside the filaments, there is very little sensitivity to the $H_0$ assumed in the reconstruction  \citep{2022A&A...657L..17K}.

The dimensionality of the integral in Equation \ref{eq:POSTERIOR} is the length of the SNIa data vector, and such high-dimensional integrals can be extremely challenging to compute, even with nested sampling \citep{nested_sampling_1, nested_sampling_2}. Attempting to compute this integral with the \codefont{dynesty} nested sampler \citep{2009MNRAS.398.1601F,2015MNRAS.450L..61H,2015MNRAS.453.4384H,dynesty,dynesty2}, led to reasonable posteriors on $H_0$, which could not, nonetheless, become robust to the sampler's tunable parameters. Further, the reported uncertainties were underestimated. Therefore, due to the challenging nature of computing this high-dimensional noisy integral, we will proceed with an analytical approach. 

We will expand $\mu$ in a linear Taylor expansion, which will yield a Gaussian posterior (Laplace approximation). In what follows, we will derive an analytically tractable form for the $H_0$ posterior, where we will approximate the integrand as a multivariate Gaussian, beginning with
\begin{equation}
    p(\alpha_\mathrm{B}|\hat{z}, \hat{\mu}, \bar{z}_\mathrm{p}) \propto\pi(\alpha_\mathrm{B})\int d z_\mathrm{c} \prod_i z_{\mathrm{c}, i}^2 \exp\left( -\frac{Q}{2}\right),
\end{equation}
where
\begin{equation}
Q = (\hat{\mu}-\bar{\mu})^\mathrm{T}\Sigma_\mathrm{\mu}^{-1}(\hat{\mu}-\bar{\mu}) + \bar{q}^\mathrm{T}\left(\Sigma+\tilde{C}\right)^{-1} \bar{q}, 
\end{equation}
and
\begin{equation}
\bar{q}_i = \hat{z}_i - (1+z_{\mathrm{c}, i})\left[1+\bar{z}_{\mathrm{p}, i}(z_{\mathrm{c}, i})\right] - 1.
\end{equation}
We will expand around the $\tilde{z}$ that minimises $\bar{q}^2$, such that
\begin{equation}
    \bar{\mu}_i(z_{\mathrm{c}, i}) \approx \bar{\mu}(\tilde{z}_i) + \mu'_i (z_{\mathrm{c}, i}-\tilde{z}_i),
\end{equation}
where $\mu'_i = \frac{\partial \mu_i}{\partial z}$ at $z=\tilde{z}_i$. We minimise $\bar{q}^2$ over a redshift grid at the lowest resolution that guarantees no new voxels are added to the line of sight to each SNIa when increasing the resolution further, in order to avoid artifacts resulting from the discrete nature of the velocity field. Therefore, we arrive at
\begin{equation}
    \hat{\mu}_i - \bar{\mu}_i \approx \mu'_i(z_{\mathrm{c}, i} - z_{0, i}),
\end{equation}
where
\begin{equation}
    z_{0, i} = \tilde{z}_i - \frac{\hat{\mu}_i}{\mu'_i} + \frac{\mu_\mathrm{c}(H_{0,\mathrm{fid}}, \tilde{z})}{\mu'_i} + \frac{10\log\left[1+\bar{z}_{\mathrm{p}}(\tilde{z}_i)\right]}{\mu'_i} - \frac{5\alpha_\mathrm{B}}{\mu'_i}.
\end{equation}
The above expansion is likely to be inadequate for photometric redshift errors, but for typical spectroscopic uncertainties a linear Taylor expansion about $\tilde{z}$ is sufficient. Setting $\Lambda = \mathrm{diag}(1/\mu')$, we arrive at
\begin{equation}
 Q = (z_\mathrm{c}-z_0)^\mathrm{T} \Lambda^\mathrm{T} \Sigma_\mathrm{\mu}^{-1} \Lambda (z_\mathrm{c}-z_0) + (z_\mathrm{c}-\tilde{z})^\mathrm{T}(\Sigma+\tilde{C})^{-1}(z_\mathrm{c}-\tilde{z}).
\end{equation}
This is a Gaussian integral in $z_\mathrm{c}$, if we approximate the $z_\mathrm{c}^2$ prior by $\tilde{z}^2$, and extend the lower integration limits to $-\infty$. Thus, we arrive at
\begin{equation}
 I \propto \frac{1}{\sqrt{|A/2\pi|}} \exp{\left(\frac{1}{2}B^\mathrm{T}A^{-1}B\right)},
\end{equation}
where
\begin{eqnarray}
    A &=& \Lambda^\mathrm{T} \Sigma_\mathrm{\mu}^{-1} \Lambda + (\Sigma + \tilde{C})^{-1}\\
    B^\mathrm{T} &=& z_0^\mathrm{T} \Lambda^\mathrm{T} \Sigma_\mathrm{\mu}^{-1} \Lambda + \tilde{z}^\mathrm{T}(\Sigma+\tilde{C})^{-1}.
\end{eqnarray}
Applying the Woodbury formula, we arrive at the simple expression
\begin{equation}
    I \propto \pi(\alpha_\mathrm{B}) \exp{\left[-\frac{1}{2}b^\mathrm{T}\tilde{B}^{-1}b\right]},
\end{equation}
where
\begin{eqnarray}
    b &=& z_1-\frac{5\alpha_\mathrm{B}}{\mu'}-\tilde{z},\\
    z_1 &=& \tilde{z}_i - \frac{\hat{\mu}_i}{\mu'_i} + \frac{\mu_\mathrm{c}(H_{0,\mathrm{fid}}, \tilde{z})}{\mu'_i} + \frac{10\log(1+\bar{z}_{\mathrm{p}, i})}{\mu'_i},\\
    \tilde{B} &=& (\Lambda^\mathrm{T} \Sigma_\mathrm{\mu}^{-1} \Lambda)^{-1}+\Sigma + \tilde{C}.
\end{eqnarray}
Finally, the integral can be simplified to
\begin{equation}
    I \propto \pi(\alpha_\mathrm{B}) \exp{\left[-\frac{\left(5\alpha_\mathrm{B}-\hat{\beta}\right)^2}{2\sigma_\beta^2}\right]},
\end{equation}
where $\hat\beta$ is the maximum a posteriori (MAP) value of $5\alpha_B$ as we assume a uniform prior
\begin{eqnarray}
    \label{eq:beta_hat}
    \hat{\beta} &=& \frac{\mathbb{I}^\mathrm{T}(\Sigma_\mathrm{\mu} + \tilde{\Sigma}_q)^{-1}\Lambda^{-1}z_2}{\mathbb{I}^\mathrm{T}(\Sigma_\mathrm{\mu} + \tilde{\Sigma}_q)^{-1}\mathbb{I}},
\end{eqnarray}
and
\begin{eqnarray}
    \tilde{\Sigma}_q &=& \Lambda^{-1}(\Sigma+\tilde{C}) \Lambda^{-1},\\
    z_2 &=& z_1 - \tilde{z},\\
    \sigma_\beta^2 &=& \left[\mathbb{I}^\mathrm{T}(\Sigma_\mathrm{\mu}+\tilde{\Sigma}_q)^{-1}\mathbb{I}\right]^{-1},
    \label{eq:gaussian_integral}
\end{eqnarray}
    and $\mathbb{I}$ is a vector of 1s.
The above may be written in an easily interpretable form
\begin{eqnarray}
\hat{\beta} &=& \frac{\Sigma_i w_i \hat{\beta}_i}{\Sigma_j w_j},\\
w &=& \mathbb{I}^\mathrm{T}(\Sigma_\mathrm{\mu} + \tilde{\Sigma}_q)^{-1},\\
\hat{\beta}_i &=& \bar{\mu}(\tilde{z}_i)-\hat{\mu}_i,
\end{eqnarray}
i.e. the maximum a posteriori value is a weighted sum of estimates $\hat{\beta}_i$ from each SNIa (with weights, $w_i$, which reduce to minimum variance weights if the velocity field is uncorrelated).

\begin{figure*}
\centering
\includegraphics[width=\textwidth]{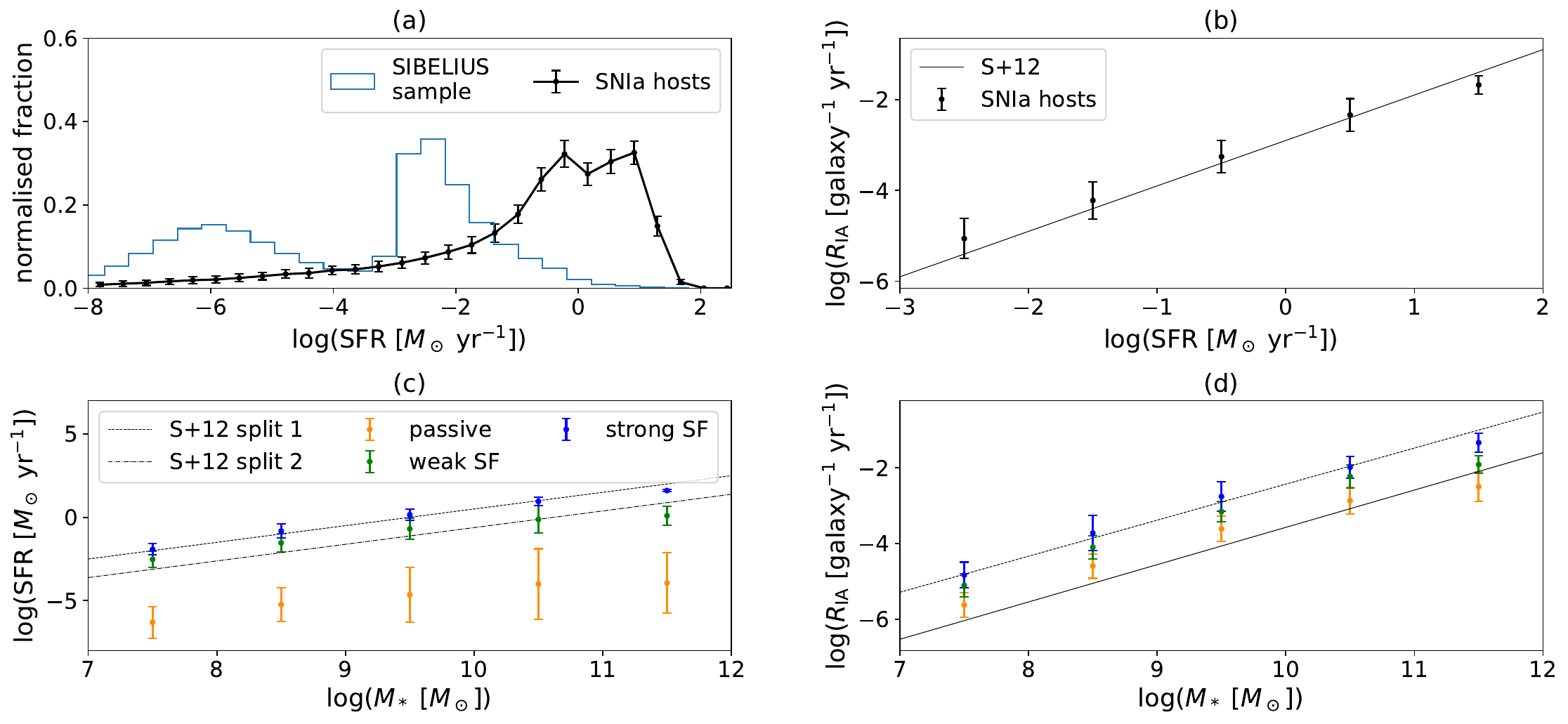}
\vspace*{-5mm}
\caption{(a) Distributions of the \codefont{SIBELIUS} SNIa host star-formation rates. In light blue, we show the star-formation rate distribution of the entire \codefont{SIBELIUS-DARK} galaxy sample. (b) SNIa rate per galaxy as a function of the star-formation rate. The continuous line is from \citet[][Figure 4]{Smith12}. (c) SNIa hosts on the star-formation rate -- stellar mass plane, coloured according to their specific star-formation rate. The black dashed line separates highly- from moderately star forming galaxies in \citet[][Figure 2]{Smith12}. The dashed dotted one encloses all weakly star-forming hosts in that plot. (d) SNIa rate per galaxy as a function of host stellar mass. The lines are from \citet[][Figure 3]{Smith12}. The error bars indicate $1\sigma$ uncertainty over the hosts in 300 Poisson realisations.}
\label{fig_7}
\end{figure*}

\subsection{Mock data generation}

In this work, we will validate the above framework on self-consistent mock data, use it to explore the impact of peculiar velocities on $H_0$ with a statistically principled treatment of observational uncertainties and demonstrate that SNeIa at $z<0.023$ need not be discarded in an $H_0$ inference, since their peculiar velocities can be accounted for. For this purpose, we first create simulated redshifts and distance moduli. The mock data generation is as follows
\begin{enumerate}
    \item Draw a Poisson realisation of SNe from Equation \ref{eq:rates}.
    \item Grid their hosts in three-dimensions using Nearest Grid Point assignment and the known cosmological redshifts, $z_\mathrm{c}^\mathrm{t}$, right ascension and declination from the \codefont{SIBELIUS-DARK} catalogue.
    \item The 2M++ non-linear peculiar velocity at those locations will be $v_\mathrm{r}$ and the corresponding peculiar redshift will be $\bar{z}_\mathrm{p}^\mathrm{t}=v_\mathrm{r}/c$.
    \item The observed peculiar redshifts are drawn from a multivariate Gaussian with mean $\bar{z}_\mathrm{p}^\mathrm{t}$ and covariance $C$. This represents our assumption that the true peculiar velocities are known within the uncertainty with which the velocity reconstruction could constrain them. The simulation velocities are $\bar{z}_\mathrm{p}^\mathrm{t}$.
    \item The observed distance moduli are drawn from a multivariate Gaussian with mean $\mu_\mathrm{c}(H_{0,\mathrm{true}},z_\mathrm{c}^\mathrm{t})+10\log{(1+\bar{z}_\mathrm{p}^\mathrm{t})}$ and covariance $\Sigma_\mathrm{\mu}$ = diag($\sigma_\mathrm{\mu}^2$).
    \item The observed redshifts are drawn from a multivariate Gaussian with mean $(1+z_\mathrm{c}^\mathrm{t})(1+\bar{z}_\mathrm{p}^\mathrm{t})-1$ and covariance $\Sigma$ =diag{($\sigma_\mathrm{z}^2$)}, where $\bar{z}_\mathrm{p}^\mathrm{t}$ is the 2M++ peculiar redshift at $z_\mathrm{c}^\mathrm{t}$.
\end{enumerate}
We assume $\sigma_\mathrm{\mu}=0.13$ mag \citep[][and references therein]{2023ApJS..264...22G}. For the redshift covariance, we assume a diagonal matrix with uncertainties $\sigma_\mathrm{z} = 0.001$ for all SNeIa. While this uncertainty can be reduced by considering host redshifts \citep{2024arXiv240611680A}, here we assume this larger scatter to demonstrate the robustness of the model to it. For the generation of mock data, we use the one 2M++ peculiar velocity realisation that has the same initial conditions as the ones with which \codefont{SIBELIUS} was constrained. For our analysis, we use the mean 2M++ peculiar velocity across all realisations. This is because only one among the 2M++ realisations will correspond to the velocity field in the Universe, but we can only infer the velocity field within observational uncertainties. Further, for the analysis of the data, we assume the 2M++ covariance matrix estimated at $\tilde{z}$, whereas for the simulated SNeIa velocities we have assumed the velocity covariance at $z_\mathrm{c}^\mathrm{t}$.

\section{Results}\label{results}

\subsection{SNIa rate modelling}

In Figure \ref{fig_7}, we show our results on the derived SNIa rates. In (a), we show the SNIa host star-formation rates at $z=0$, in comparison to those of the entire galaxy sample. The latter presents a strong bimodality corresponding to passive and star-forming galaxies. The bimodality is seen in observations of galaxy colours \citep{2019MNRAS.488.3929C}. As shown in low-redshift SNIa observations in \citet[][Figure 13]{2013ApJ...770..107C}, the distribution peaks in the range $-1< \log_{10}$(SFR/[$M_\odot$ yr$^{-1}$])$<1$, in accordance with our results. The error bars represent the uncertainty from 300 Poisson realisations. In (b), we plot the SNIa rate in star-formation rate bins. We compare with the corresponding plot from the low-redshift SDSS-II Supernova Survey \citep[][Figure 4]{Smith12} (`S+12'). Our rates are consistent with the linear fit. In (c), we plot the SNIa hosts on the SFR-$M_*$ plane. We consider a host to be passive if $\log_{10}$(sSFR/yr$^{-1}$)$<-12$, weakly star-forming if $-12<\log_{10}$(sSFR/yr$^{-1}$)$<-9.5$, and strongly star-forming if $\log_{10}$(sSFR/yr$^{-1}$)$>-9.5$. The SNIa hosts follow the same trends as \citet[][Figure 2]{Smith12}. In Figure \ref{fig_7}c and d, we add the `split 1' line to visually envelope the distribution of strongly star-forming hosts in the above study. This is the dashed line in \citet[][Figure 2]{Smith12}. `split 2' represents the dashed-dotted line in the above figure. Note that \citet{Smith12} assigned random SFRs to passive hosts, whereas we maintain the values by \codefont{GALFORM}. Finally, in (d) we show the SNIa rates as a function of stellar mass, compared to \citet[][Figure 3]{Smith12}. The overall trends agree within error bars, with the exception of the highest-mass bin of strongly star-forming galaxies in (c), which contains very few members for a statistically meaningful report on the moments of the SFR distribution. We therefore consider the rates to be sufficiently realistic for the purposes of our analysis here, as we will show in what follows.

\begin{figure*}
\centering
\includegraphics[width=\textwidth]{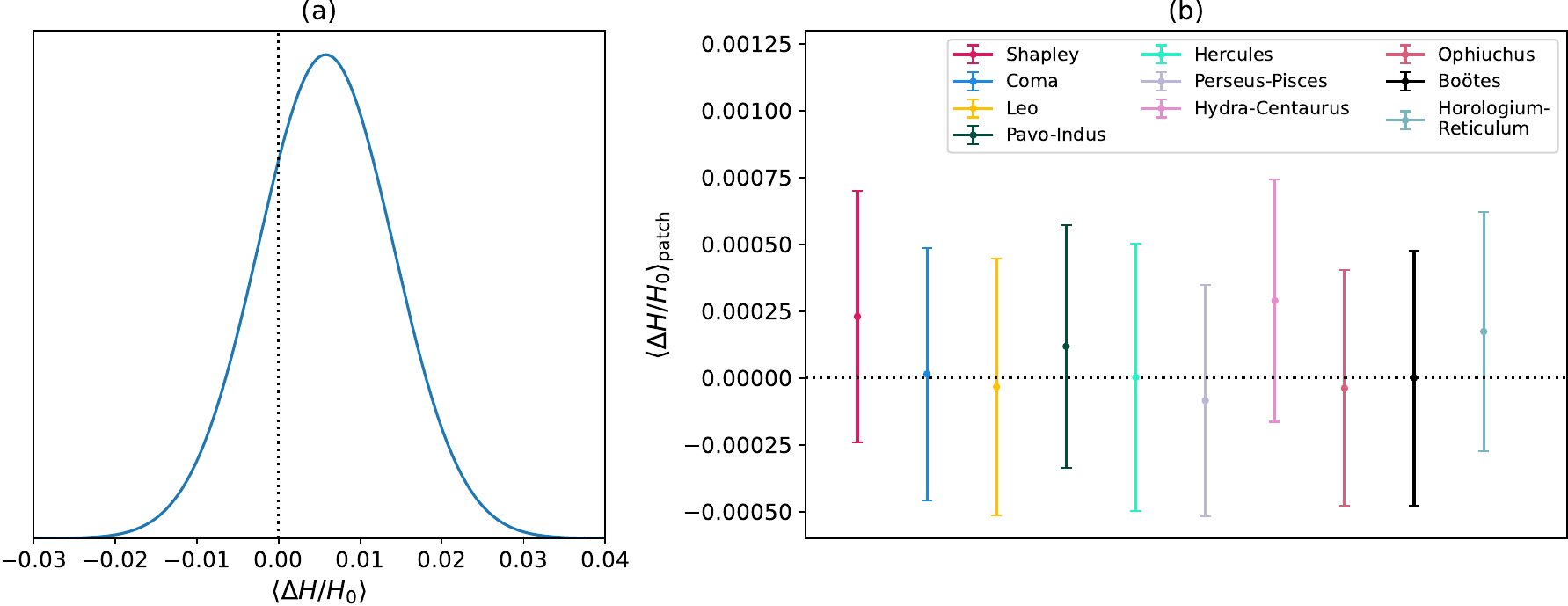}
\vspace{-5mm}
\caption{(a) $\Delta H/H_0$ posterior averaged over 100 SNIa realisations with 100 noise realisations each when peculiar velocities are ignored in the $H_0$ inference. We find on average $\langle \Delta H/H_0 \rangle = 0.006\pm0.008$ ($0.4 \pm 0.5$ km s$^{-1}$ Mpc$^{-1}$) in the local Universe in the range $0.023<z<0.046$ assuming $\sigma_\mathrm{\mu} = 0.13$ mag, $\sigma_z = 0.001$ and $N_\mathrm{SN} \sim $ 200 SNe in each dataset. The latter is determined by the number of available 2M++ realisations. (b) Contribution to $H_0$ variation in the direction of known structures in the Universe. The Hydra - Centaurus and Shapley superclusters contribute the most to the increase in $H_0$ when peculiar velocities are ignored, but the evidence is very weak.}
\label{fig_1}
\end{figure*}

\subsection{Bayesian hierarchical framework}

For the first part of our analysis, we investigate the effect of the specific configuration of the non-linear peculiar velocities in our local Universe on $H_0$. For this purpose, we generate mock data with non-linear peculiar velocities added and wrongly ignore them in our analysis. We run the analysis across $\sim 100$ mock datasets of $\sim 200$ SNeIa each at $0.023<z<0.046$ and $\sim 100$ noise realisations for each dataset, beyond which our results do not change significantly. This is because we want to probe the effect of the local Universe on the average SNIa sample. We present our results in Figure \ref{fig_1}. We find that, on average, the configuration of the local large-scale structure gives rise to $\langle\Delta H/H_0\rangle = 0.006 \pm 0.008$ ($0.4 \pm 0.5$ km s$^{-1}$ Mpc$^{-1}$) in the presence of the observational uncertainties assumed here. 

In the second part of our analysis, we probe what drives this mild positive change on $H_0$. Equation \ref{eq:beta_hat} (along with its associated weight) provides a proxy for the contribution of each source to the $H_0$ posterior for each dataset. We therefore derive $\beta_i$ and $w_i$ for each source in the mock datasets. We isolate all sources at $\pm 10^{\circ}$ in right ascension and declination from known structures in the local Universe and report the average effect $\langle \Delta H/H_0 \rangle_\mathrm{patch} = \sum_i w_i \beta_i/(5\sum_j w_j)$, where $i$ runs over the sources in each structure and $j$ runs over all sources in the sample. We choose the querying radius of $\pm 10^{\circ}$ to minimise overlap of the patches. This configuration yields $(4.2, 1.43, 1.75, 1.45, 3.34, 1.94, 2.65, 0.93, 1.42, 0.71)$ sources per patch in the Shapley, Coma, Leo, Pavo-Indus, Hercules, Perseus-Pisces, Hydra-Centaurus, Ophiuchus, Boötes and Horologium-Reticum superclusters, respectively. The number of SNe in each dataset is limited by the number of available velocity fields, preventing tighter error bars. Upcoming reconstructions will provide more velocity fields with more advanced velocity modelling, allowing a more precise and accurate determination of the effect of local superstructures individually. We find that the largest positive contribution to $H_0$ at $0.023<z<0.046$ is from SNeIa in the direction of the Hydra-Centaurus and Shapley superclusters, although the evidence is weak. SNeIa which on average exhibit $\langle \Delta H/H_0 \rangle_\mathrm{patch}>0$ have predominantly positive peculiar velocities, whereas those associated with $\langle \Delta H/H_0 \rangle_\mathrm{patch}<0$ have mainly negative peculiar velocities. Given the diminishing impact of peculiar velocities on $H_0$, the above results are consistent with the value reported by \citet{odderskov} who placed the impact of non-linear velocities in random N-body simulations at $\sim 0.3\%$ on $H_0$ in the wider redshift range $0.01<z<0.1$, ignoring velocity correlations.

\cite{odderskov} explored the impact of peculiar velocities on $H_0$ in random simulations assuming the SNIa redshift distribution of an observed sample and one resulting from rates proportional to the halo mass, finding that the rate choice affected the derived $H_0$. To test this statement in a constrained simulation and assess the sensitivity of the first part of our analysis to the SNIa rate modelling, we repeat our analysis assuming uniform rates across the simulated galaxies instead of the rates in Equation \ref{eq:rates}. Assuming uniform rates, we find $\langle \Delta H/H_0 \rangle = 0.006 \pm 0.008$, i.e. no change on average with respect to the results where we derived SNIa rates from the galaxy star-formation histories. We therefore conclude that the per-host SNIa rate modelling is unlikely to significantly change our conclusions for a typical SNIa sample, as it does not significantly impact the host velocity and distance distributions at the level of the uncertainties assumed here. The latter is also corroborated by the fact that the 1-point velocity statistics show no evidence of relative SNIa host velocity bias with respect to galaxies. Our findings suggest that it is the location of SNIa hosts in the large-scale structure which predominantly drives the derived $\langle \Delta H/H_0 \rangle >0$, rather than the preferential occurrence of SNeIa in star-forming hosts which occupy specific velocity environments. Therefore, once a constrained simulation of a cosmological volume has been obtained, accurate SNIa rate modelling, which is often challenging to perform, may be of secondary importance to the study of velocity dynamics. Further work is required at the level of 2-point statistics and beyond to probe the SNIa velocity bias.

\begin{figure*}
\centering
\includegraphics[width=\textwidth]{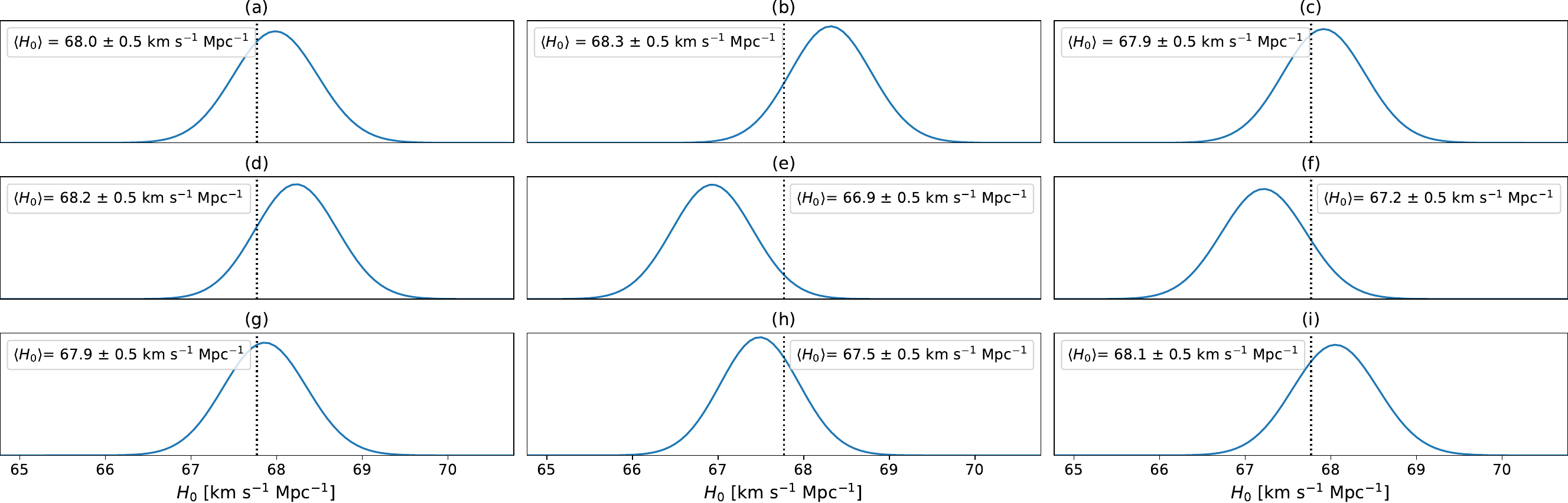}
\vspace{-5mm}
\caption{$H_0$ posterior from the Bayesian hierarchical model in in Equation \ref{eq:POSTERIOR} in the presence of observational uncertainties for 9 random SNIa datasets at $z<0.023$, after wrongly assuming $H_0=63$ km s$^{-1}$ Mpc$^{-1}$ when the data have been generated with $H_{0,\mathrm{true}} = 67.77$ km s$^{-1}$ Mpc$^{-1}$. The dotted line represents the true $H_{0,\mathrm{true}}$ in the \codefont{SIBELIUS} simulation. The datasets contain $N_\mathrm{SN} \sim$ 165 SNeIa each, determined by the number of available 2M++ realisations and sources in the ZTF Bright Transient Survey Sample Explorer.}
\label{fig_9}
\end{figure*}

The statistical framework for $H_0$ inference presented here can be used to extend the SNIa redshift range to $z<0.023$, which are typically discarded, as the effects of peculiar velocities are larger. To demonstrate this, we generate data with $H_0=H_{0,\mathrm{true}}=67.77$ km s$^{-1}$ Mpc$^{-1}$, wrongly assume in our analysis that $H_0^\mathrm{fid}=63$ km s$^{-1}$ Mpc$^{-1}$ and infer $H_0$. The choice of $H_0^\mathrm{fid}$ is arbitrary and our results do not depend on it, because we infer the offset of $H_0$ from the truth. In Figure \ref{fig_9}, we show the $H_0$ posterior for 9 random datasets with $\sim 165$ SNeIa each, according to the current number of sources in the Zwicky Transient Facility (ZTF) Bright Transient Survey (BTS) Sample Explorer \citep{2020ApJ...895...32F,2020ApJ...904...35P}. This was the number of SNe in the BTS sample explorer with the appropriate cuts at $z<0.023$. A similar number of SNeIa suitable for our analysis is available from the Pantheon+SH0ES dataset \citep{2022ApJ...938..113S,R22}. We show only 9 mocks for the purposes of presentation, but 100 were checked and found consistent with the ground truth. The posteriors are consistent with $H_{0,\mathrm{true}}$. The posteriors are also consistent with the ground truth when we set $H_0^\mathrm{fid}>H_{0,\mathrm{true}}$. The consistency persists even when we generate data with the \codefont{SIBELIUS} peculiar velocities, which have been generated by a full gravity solver which potentially produces more non-linear velocities than the \codefont{BORG} quasi-linear velocity model. The \codefont{SIBELIUS} velocities are at the scale of individual hosts, so they, in principle, account for velocity dispersion to some degree. However, we present our results using data generated with the 2M++ velocities, as only the velocity covariance matrix and mean velocity field from the 2M++ inference are available.

\section{Discussion \& conclusions}\label{conclusions}

In this work we presented a more complete treatment of peculiar velocities for $H_0$ inference at the level of the three-dimensional field, marginalising over the SNeIa's unknown cosmological redshifts and peculiar velocities. A number of studies have corrected the observed redshifts for peculiar velocities \citep[e.g.][]{2022PASA...39...46C}, and the correction is uncertain and should be marginalised over \citep[Equation 10,][]{2022ApJ...938..112P}. In our approach, all assumed contributions to the observed redshifts (peculiar velocities and errors, redshift errors) are forward-modelled and accounted for in the distance modulus likelihood. Further, instead of assuming a typical dispersion of $250$ km/s for all sources, we utilise our knowledge of the constrained non-linear velocity field at $2.65$ Mpc/h resolution to provide improved velocity uncertainties and their correlations, accounting for the varying quality of the reconstruction at each location \citep{2022PASA...39...46C}. The overall difference can be understood by comparing our Equation \ref{eq:reordering} to \citet[Equation 10, ][]{2022ApJ...938..112P}. Instead of each SNIa being associated with a peculiar velocity estimate, it is associated with a full velocity posterior along its line of sight, which accounts for the quality of the reconstruction, redshift errors and non-linear velocity correlations. All these uncertainties are part of the Bayesian modelling of the distance modulus likelihood.

Comparing the simulated SNIa host galaxy properties to observations in a similar redshift range recovers good agreement up to a mild rescaling factor, which could originate in the semi-analytic nature of \codefont{GALFORM} and to which our analysis is robust. We found an insignificant increase of $\sim 0.4 \pm 0.5$ km s$^{-1}$ Mpc$^{-1}$ on $H_0$ on average in the range $0.023<z<0.046$. The upper redshift cut is imposed by the redshift extent of the \codefont{SIBELIUS-DARK} simulation. This increase is weak, predominantly in the direction of the Shapley and Hydra - Centaurus superclusters. Given the diminishing impact of peculiar velocities on $H_0$ at higher redshifts, we conclude that it is unlikely that the origin of the Hubble tension is in the assumptions on the velocity dynamics or the specific configuration of our local Universe. This conclusion is reinforced by \citet{2017MNRAS.471.4946W}, who found a sample variance error on $H_0$ at 0.3 km s$^{-1}$ Mpc$^{-1}$ in the range $0.023<z<0.15$ using unconstrained simulations and observers hosted by haloes of varying properties.

Repeating the analysis assuming uniform SNIa rates instead of rates that depend on the star-formation history makes no significant difference to our main conclusions. This suggests that the per-host SNIa rate modelling is of secondary importance to the study of SNIa velocity dynamics, indicating that the positive $\Delta H/H_0$ is driven by the locations of galaxies in the large-scale structure, rather than the particular star-forming locations of SNeIa. These results further indicate that once a constrained simulation of a cosmological volume is available, precise modelling of the SNIa rate -- often challenging -- could be less critical for studying velocity dynamics at SNIa locations. The $H_0$ errors reported here are likely expected to increase further in the case of a real-data application, where one might want to include the effects of velocity dispersion and peculiar velocities sourced outside the 2M++ volume, but already broadly agree in magnitude with the inference of the impact of linear velocities on $H_0$ from ZTF DR2 data \citep{carreres2024ztf}. This implies that the inclusion of non-linear velocity correlations leaves the conclusions on the impact of velocities on $H_0$ qualitatively invariant for a typical SNIa sample with respect to the \citet{carreres2024ztf} analysis, which assumed a linear velocity covariance in the range $0.023<z<0.06$. However, for the purposes of an accurate quantitative estimation or inferring $H_0$ from SNeIa within our local flow ($z<0.023$), accounting for non-linear velocity correlations is necessary. This is because at $z<0.023$ where contributions to the observed redshifts from peculiar velocities are significant, uncertainties which are not accounted for in the $H_0$ posterior self-consistently may, in principle, introduce biases. Our method accounts both for non-linear velocity correlations and the 2M++ survey uncertainties self-consistently, combined with state-of-the-art velocity modelling \citep{2025arXiv250200121S}. It remains to be checked against real data whether the more accurate velocity modelling in the \codefont{BORG} 2M++ reconstruction will yield results on $H_0$ at $z<0.023$ which will be consistent with existing studies. Therefore, the effectiveness of the Bayesian treatment to deal with peculiar velocity effects allows the use, in conjunction with posterior samples of the local peculiar velocity field, of low-redshift SNe that are normally discarded from Hubble constant analyses. The uncertainties in the presented $H_0$ posteriors are dominated by the number of SNeIa used, so extending the minimum redshift of $H_0$ analyses to $z\sim0$ will reduce the variance of the $H_0$ posteriors presented here.

\section*{Acknowledgements}\label{acknowledgements}
We thank the anonymous referee for the useful comments, which improved the quality of the manuscript. We thank Stuart McAlpine for the \codefont{SIBELIUS-DARK} catalogue, Jens Jasche and Guilhem Lavaux for the 2M++ \codefont{BORG} inference data and their comments on the draft. We thank Josh Speagle for his insights on the performance of \codefont{dynesty} in high dimensions. ET would like to thank Ariel Goobar for his comments on the draft, Bruce Bassett, Boris Leistedt and Michelle Lochner for useful discussions. This work was supported by STFC through Imperial College Astrophysics Consolidated Grant ST/W000989/1. This research utilised the HPC facility supported by the Research Computing Service at Imperial College London. ET further acknowledges support from the Centre for Cosmological Studies Balzan Fellowship, that contributed to the successful completion of this work. This work has been done within the Aquila Consortium (\url{https://www.aquila-consortium.org}).

\section*{Data availability}

Data products underlying this article can be made available upon reasonable request to the corresponding author. 



\bibliographystyle{mnras}
\bibliography{example} 




\bsp	
\label{lastpage}
\end{document}